\documentclass[twocolumn,showpacs,preprintnumbers,amsmath,amssymb]{revtex4}
\usepackage{graphicx}% Include figure files
\usepackage{dcolumn}% Align table columns on decimal point
\usepackage{bm}% bold math

\begin{document}

\preprint{APS/123-QED}

\title{Supernova Neutrino Nucleosynthesis of Light Elements \\
with Neutrino Oscillations}% Force line breaks with \\

\author{Takashi Yoshida $^1$}\email{takashi.yoshida@nao.ac.jp}
\author{Toshitaka Kajino $^{2,3}$}
\author{Hidekazu Yokomakura $^4$}
\author{Keiichi Kimura $^4$}
\author{Akira Takamura $^5$}
\author{Dieter H. Hartmann $^{6}$}
\affiliation{%
$^1$Astronomical Institute, Graduate School of Science, Tohoku University, 
Miyagi 980-8578, Japan \\
$^2$National Astronomical Observatory of Japan, and The Graduate University 
for Advanced Studies, Tokyo 181-8588, Japan \\
$^3$Department of Astronomy, Graduate School of Science, University of Tokyo, 
Tokyo 113-0033, Japan \\
$^4$Department of Physics, Graduate School Science, Nagoya University, 
Aichi 464-8602, Japan \\
$^5$Department of Mathematics, Toyota National College of Technology, 
Aichi 471-8525, Japan \\
$^6$ Department of Physics and Astronomy, Clemson University, Clemson,
South Carolina 29634, USA
}%

\date{\today}% It is always \today, today,

\begin{abstract}
Light element synthesis in supernovae through neutrino-nucleus
interactions, i.e., the $\nu$-process, is affected by neutrino oscillations
in the supernova environment. There is a resonance of 13-mixing in the O/C 
layer, which increases the rates of charged-current $\nu$-process reactions 
in the outer He-rich layer.
The yields of $^7$Li and $^{11}$B increase by about a factor of 1.9 and 1.3,
respectively, for a normal mass hierarchy and an adiabatic 13-mixing resonance, 
compared to those without neutrino oscillations.
In the case of an inverted mass hierarchy and a non-adiabatic 13-mixing resonance,
the increase in the $^7$Li and $^{11}$B yields is much smaller.
Observations of the $^7$Li/$^{11}$B ratio in stars showing
signs of supernova enrichment could thus provide a unique test of 
neutrino oscillations and constrain their parameters and the mass hierarchy.
\end{abstract}

%26.30.+k Nucleosynthesis in novae, supernovae, and other explosive 
%         environments
%14.60.Pq Neutrino mass and mixing
%97.60.Bw Supernovae
%25.30.Pt Neutrino scattering
%13.15.+g Neutrino interactions

\maketitle

A tremendous number of neutrinos are released from a core-collapse 
supernova (SN).
These neutrinos interact with nuclei in the surrounding stellar envelope and 
thereby affect the synthesis of new elements.
This so-called $\nu$-process may be a major contributor to the production of 
several light isotopes, such as $^7$Li, $^{11}$B, $^{19}$F, 
as well as a few heavy isotopes, such as $^{138}$La and $^{180}$Ta
\cite{wh90,ww95,ga01,ra02,yt04,hk05,yk05}.
However, the yields of these isotopes may depend on the effects of neutrino 
oscillations, which was not taken into consideration in the above cited studies.

Recent neutrino experiments on atmospheric \cite{SK04}, solar \cite{SNO04},
and reactor neutrinos \cite{ap03,KL05} significantly constrain most of the 
neutrino oscillation parameters.
However, only an upper limit on $\theta_{13}$ is obtained \cite{ap03}
and the mass hierarchy remains unknown. Theoretical studies of neutrino 
oscillations in SNe have been used to suggest potential constraints on 
$\theta_{13}$ and the mass hierarchy based on observed SN neutrino spectra. 
These studies indicate that the neutrino spectra from SNe strongly depend on 
$\theta_{13}$ and the assumed mass hierarchy \cite{ds00,ha04}.
When the resonance of the 13-mixing is adiabatic, substantial conversion  
$\nu_{\rm e} \leftrightarrow \nu_{\mu,\tau}$ occurs in the O/C layer for a 
normal mass hierarchy and conversion 
$\bar{\nu}_{\rm e} \leftrightarrow \bar{\nu}_{\mu,\tau}$ occurs
for an inverted hierarchy. These direct methods are of course limited by the 
fact that nearby core-collapse SNe occur rarely in the small detection volume 
given by current detector sizes and methods.

Here we suggest an alternative method to study the effects of neutrino 
oscillations, by considering light element synthesis in SNe. 
Neutrino energy spectra change as they are transported through SN ejecta 
\cite{ds00}.
This change will affect the production of light elements via the $\nu$-process.
The thermal neutrinos emitted from a cooling protoneutron star have a 
well-known, but not yet rigorously established, energy hierarchy;
$\langle \varepsilon_{\nu_{\rm e}} \rangle <
\langle \varepsilon_{\bar{\nu}_{\rm e}} \rangle <
\langle \varepsilon_{\nu_{\mu,\tau},\bar{\nu}_{\mu,\tau}} \rangle 
%= \langle \varepsilon_{\bar{\nu}_{\mu,\tau}} \rangle
$ (e.g., \cite{kr03}).
Neutrino oscillations could thus increase the average energies of $\nu_{\rm e}$
and $\bar{\nu}_{\rm e}$, and consequently the rates of charged-current
$\nu$-process reactions could be much larger than expected from models without
oscillations. Therefore, the yields of the light elements may increase significantly.

We investigate nucleosynthesis of light elements 
$^7$Li and $^{11}$B through the $\nu$-process in SNe taking neutrino oscillations
into account. Since the other $\nu$-process elements are mainly produced in the O-rich
layers \cite{ga01,hk05}, they are not expected to be affected by neutrino oscillations.
The $^7$Li and $^{11}$B yields in SNe can thus be used as probes of neutrino 
oscillations.
%We propose that the dependence of $^7$Li and $^{11}$B yields 
We show that the $^7$Li yield increases significantly through neutrino oscillations.
The dependence of the $^7$Li/$^{11}$B ratio on the mixing parameter, $\theta_{13}$, 
provides an observable signature that could be used to constrain its absolute value and 
the neutrino mass hierarchy.

Neutrino luminosities are assumed to decrease exponentially in time, 
with a decay time scale of $\sim$ 3 s \cite{wh90,ww95,ra02,yt04,hk05,yk05}.
The total neutrino energy is assumed to be fixed at $3 \times 10^{53}$ erg.
The neutrino energy spectra at the neutrino sphere are approximated 
with Fermi-Dirac (FD) distributions with zero chemical potential.
The neutrino temperatures of $\nu_{\rm e}$, $\bar{\nu}_{\rm e}$, and
($\nu_{\mu,\tau}$ and $\bar{\nu}_{\mu,\tau}$) are set to be 3.2, 
5.0, and 6.0 MeV as adopted in \cite{yt04}.
These energy spectra change during the subsequent passage through
the outer stellar layers by neutrino oscillations.

In order to evaluate the reaction rates of the $\nu$-process, we need the
cross sections as functions of neutrino energy because the spectra changed 
by neutrino oscillations no longer follow the FD shape.
We assume that the cross sections of the charged-current reactions
of $^4$He and $^{12}$C, i.e., $^4$He($\nu_{\rm e},{\rm e^-}p)^3$He,
$^4$He($\bar{\nu}_{\rm e},{\rm e^+}n)^3$H,
$^{12}$C($\nu_{\rm e},{\rm e^-}p)^{11}$C, 
$^{12}$C($\bar{\nu}_{\rm e},{\rm e^+}n)^{11}$B,
$^{12}$C($\nu_{\rm e},{\rm e^-}\gamma)^{12}$N, and
$^{12}$C($\bar{\nu}_{\rm e},{\rm e^+}\gamma)^{12}$B, are expressed as a
power law $\sigma(\varepsilon_\nu) = 
\sigma_0 (\varepsilon_\nu - \varepsilon_{\rm th})^\alpha$,
where $\varepsilon_{\rm th}$ is the threshold energy.
Coefficients of the functions are determined such that the reaction rates
deduced using these cross sections (and assuming FD energy distributions) fit  
the rates tabulated in \cite{hw92}.
Details are provided in \cite{yk052}.
For the other $\nu$-process reactions, we use the reaction rates with FD 
distribution of the neutrino spectra.

Recent neutrino experiments \cite{SK04,SNO04,ap03,KL05} have 
determined most of the values of the mass squared differences 
$\Delta m^2_{ij} \equiv m^2_i - m^2_j$ and the mixing angles $\theta_{ij}$.
Based on these results, we use  
$\Delta m^2_{21} = 7.9 \times 10^{-5}$ eV$^2$, 
$\Delta m^2_{31} = \pm 2.4 \times 10^{-3}$ eV$^2$, and
$\sin^{2}2\theta_{12} = 0.816$, $\sin^{2}2\theta_{23} = 1.0$,
$0 \le \sin^{2}2\theta_{13} \le 1 \times 10^{-1}$.
The positive value of $\Delta m^2_{31}$ corresponds to 
^^ ^^ normal hierarchy'', i.e., $m_1 < m_2 < m_3$ and the negative value
corresponds to ^^ ^^ inverted hierarchy'', i.e., $m_3 < m_1 < m_2$.
We numerically solve the mixing probabilities of neutrinos for each
neutrino energy by Runge-Kutta methods and using the exact solutions
of the oscillations described in \cite{kt02}.
By convolving the mixing probabilities and the neutrino spectra at the
neutrino sphere, we evaluate the neutrino energy spectra taking
neutrino oscillations into account. We do not include {\it CP} phase $\delta$.
Based on \cite{kp87} we assume that the effect of {\it CP} violation will not
be seen because the spectra of $\nu_\mu$ ($\bar{\nu}_\mu$) and 
$\nu_\tau$ ($\bar{\nu}_\tau$) emitted from the neutrino sphere are the same.
The change of the spectra due to oscillations
is calculated using the density profile of a presupernova model.

\begin{figure}[t]
\includegraphics[width=7cm]{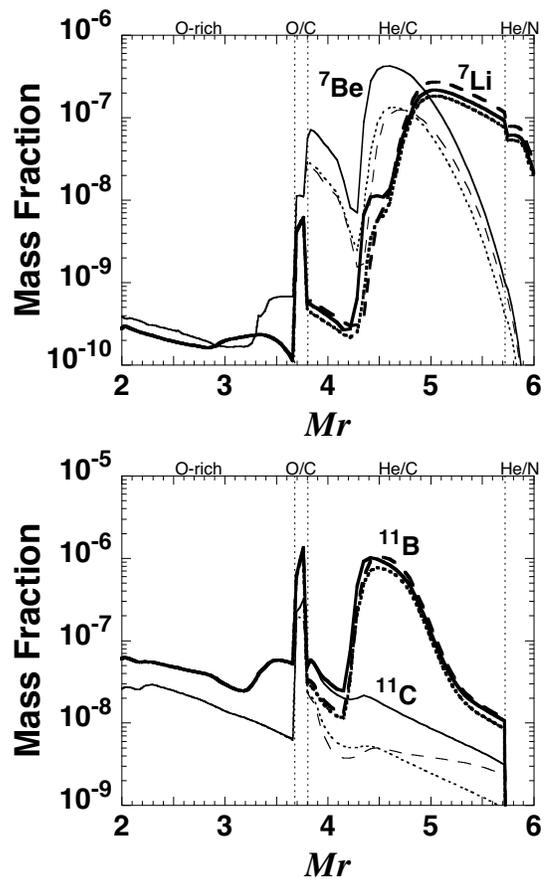}
\caption{
The mass fraction distributions of $^7$Li and its isobar $^7$Be (upper panel), 
and $^{11}$B and $^{11}$C (lower panel) in the case of 
$\sin^{2}2\theta_{13}=2 \times 10^{-2}$.
Thick lines indicate the distributions of $^7$Li and $^{11}$B.
Thin lines indicate the distributions of $^7$Be and $^{11}$C.
Solid lines and dashed lines correspond to a normal hierarchy and
inverted hierarchy, respectively.
Dotted lines correspond to the case without neutrino oscillations.
The horizontal axis is the interior mass in units of the solar mass.
}
\end{figure}

We use the same SN explosion model as in \cite{yt04,yk05}.
The presupernova model is the 14E1 model constructed for SN 1987A 
in \cite{sn90}.
The SN explosion is calculated using piecewise parabolic method
code \cite{cw84,sn92}.
The explosion energy and the location of the mass cut are set to be
$1 \times 10^{51}$ erg and 1.61 $M_\odot$.
The detailed nucleosynthesis in the SN is calculated using a
nuclear reaction network including 291 species 
\cite{yt04}.

Figure 1 shows the mass fraction distributions of $^7$Li and $^{11}$B
in the SN ejecta with neutrino oscillations of 
$\sin^{2}2\theta_{13} = 2 \times 10^{-2}$ and for those without oscillations.
%In this figure, the mass fractions of $^7$Li and its isobar $^7$Be are
%shown separately.
%The mass fractions of $^{11}$B and $^{11}$C are also separated.
In the case of a normal hierarchy, the mass fraction of $^7$Be with the
neutrino oscillations is much larger than that without oscillations
in the He layer.
There is a 13-mixing resonance for neutrinos in the O/C layer and 
the resonance is adiabatic in this case.
Thus, the energy spectrum of $\nu_{\rm e}$ in the He/C layer becomes almost 
the same as  that of $\nu_{\mu,\tau}$ in the O-rich layer.
%As a result, the average $\nu_{\rm e}$ energy increases.
Beryllium 7 is produced through $^4$He($\nu,\nu'n)^3$He($\alpha,\gamma)^7$Be.
Owing to the neutrino oscillations, the reaction rate of 
$^4$He($\nu_{\rm e},{\rm e^-}p)^3$He becomes larger than that of
$^4$He($\nu,\nu'n)^3$He.
The mass fraction of $^7$Li including neutrino oscillations is also larger,
but the increment is much smaller than that for $^7$Be.
The main production process of $^7$Li is 
$^4$He($\nu,\nu'p)^3$H($\alpha,\gamma)^7$Li and the corresponding 
charged-current reaction is $^4$He($\bar{\nu}_{\rm e},{\rm e^+}n)^3$H.
However, there are no resonances for antineutrinos.

\begin{figure}[b]
\includegraphics[width=7cm]{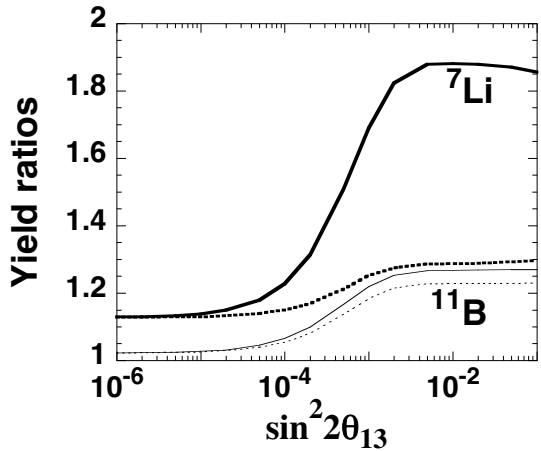}
\caption{
The yield ratios of $^7$Li and $^{11}$B with the relation of 
$\sin^{2}2\theta_{13}$.
Thick solid line and thick dotted line are the yield ratio of $^7$Li
in the cases of a normal hierarchy and inverted hierarchy, respectively.
The thin solid line and thin dotted line are that of $^{11}$B in the 
cases of normal and inverted hierarchy.
The case of $\sin^{2}2\theta_{13}=0$ is also calculated (see text).
}
\end{figure}

The effect of neutrino oscillations on the mass fraction distributions of
$^{11}$B and $^{11}$C is similar to that for $^7$Li and $^7$Be.
The mass fraction of $^{11}$C with the neutrino oscillations is larger than
that without oscillations in the He layer.
During the $\nu$-process, $^{11}$C is produced through 
$^{12}$C($\nu,\nu'n)^{11}$C and partly through
$^{12}$C($\nu_{\rm e},{\rm e^-}p)^{11}$C.
The reaction rate of $^{12}$C($\nu_{\rm e},{\rm e^-}p)^{11}$C 
with the oscillations becomes larger than that 
without oscillations by about one order of magnitude.
The mass fraction of $^{11}$B with oscillations is only slightly
larger than that without oscillations.
The main production process of $^{11}$B is
$^4$He($\nu,\nu'p)^3$H($\alpha,\gamma)^7$Li($\alpha,\gamma)^{11}$B.
The corresponding charged-current reaction is
$^4$He($\bar{\nu}_{\rm e},{\rm e^+}n)^3$H.
About $12 \% - 16 \%$ of $^{11}$B in the He layer is produced from $^{12}$C
through $^{12}$C($\nu,\nu'p)^{11}$B and 
$^{12}$C($\bar{\nu}_{\rm e},{\rm e^+}n)^{11}$B.
The $^{11}$B abundant region in the He layer is inside the $^7$Li abundant 
region because of the decrease in peak shock temperature as one moves 
outward in the star.
%The production process of $^{11}$B is mainly 
%$^4$He($\nu,\nu'p)^3$H($\alpha,\gamma)^7$Li($\alpha,\gamma)^{11}$B
%and partly $^{12}$C($\nu,\nu'p)^{11}$B.
%The corresponding charged-current reactions are
%$^4$He($\bar{\nu}_{\rm e},{\rm e^+}n)^3$H and 
%$^{12}$C($\bar{\nu}_{\rm e},{\rm e^+}n)^{11}$B.
The increase in the $^{11}$B production through
$^{12}$C($\bar{\nu}_{\rm e},{\rm e^+}n)^{11}$B is not as large because 
of the absence of resonances for antineutrinos, as mentioned above.
In the O-rich layers, light element production is not
influenced by the neutrino oscillations.
The oscillation amplitude in these layers is too small because of high
densities (e.g., \cite{ds00}).

In the case of an inverted hierarchy, mass fractions of $^7$Li and $^{11}$B
are larger than those for a normal hierarchy. The reaction rates of 
$^4$He($\bar{\nu}_{\rm e},{\rm e^+}n)^3$H and
$^{12}$C($\bar{\nu}_{\rm e},{\rm e^+}n)^{11}$B become larger owing to
an adiabatic resonance of 
$\bar{\nu}_{\rm e} \leftrightarrow \bar{\nu}_{\mu,\tau}$.
However, the increment of the mass fractions of $^7$Li and $^{11}$B is 
less pronounced than that of $^7$Be and $^{11}$C for a normal hierarchy.
This is because the average energy of $\bar{\nu}_{\rm e}$ is larger than
$\nu_{\rm e}$ at the neutrino sphere and, therefore the difference from
the average $\nu_{\mu,\tau}$ ($\bar{\nu}_{\mu,\tau}$) energy is smaller.
On the other hand, the mass fractions of $^7$Be and $^{11}$C are slightly
larger in the mass range $M_r \ge 4.5 M_\odot$ and slightly smaller inside 
the range of the He layer.
There is no 13-mixing resonance for $\nu_{\rm e}$, so that substantial 
conversion of $\nu_{\rm e} \leftrightarrow \nu_{\mu,\tau}$ does not occur.
At the same time, some $^7$Be and $^{11}$C capture neutrons produced 
through $^4$He($\bar{\nu}_{\rm e},{\rm e^+}n)^3$H.
% in the range of 
%$3.8 M_\odot \le M_r \le 4.5 M_\odot$.
%In O-rich layers, there are no effects due to neutrino oscillations
%similar to the case of a normal hierarchy.

Figure 2 shows the ratios of the $^7$Li and $^{11}$B yields with neutrino
oscillations in comparison to those without oscillations, hereafter called
yield ratios, as a function of $\sin^{2}2\theta_{13}$.
The yields of $^7$Li and $^{11}$B without the oscillations are
$2.36 \times 10^{-7} M_\odot$ and $6.26 \times 10^{-7} M_\odot$.
The yield ratio of $^7$Li is at most 1.88 in the case of 
$\sin^{2}2\theta_{13} \ge 2 \times 10^{-3}$ and normal hierarchy.
This increase in the yield is due to the adiabatic 13-mixing resonance.
%The contribution from $^7$Be increases through enhanced
%$^4$He($\nu_{\rm e},{\rm e^-}p)^3$He rate owing to large
%$\nu_{\rm e} \leftrightarrow \nu_{\mu,\tau}$ conversion.
In the case of 
$2 \times 10^{-5} \le \sin^{2}2\theta_{13} \le 2 \times 10^{-3}$,
the yield ratio of $^7$Li increases with $\sin^{2}2\theta_{13}$.
In this $\theta_{13}$ range, the 13-mixing resonance changes from non-adiabatic
to adiabatic with increasing in $\sin^{2}2\theta_{13}$.
In the case of $\sin^{2}2\theta_{13} < 2 \times 10^{-5}$, corresponding to
nonadiabatic resonance, the yield of $^7$Li is about 1.13.
%The resonance is nonadiabatic and the production
%of $^7$Be increases only in an outer range of the He layer.
In the case of an inverted hierarchy, the dependence of the $^7$Li yield on
$\sin^{2}2\theta_{13}$ is similar to the normal hierarchy case.
However, the increment of the yield ratio is much smaller.
The smaller difference of the average energy between $\bar{\nu}_{\rm e}$
and $\bar{\nu}_{\mu,\tau}$ reflects the smaller increase in the yield ratio.

The dependence of the $^{11}$B yield ratio on $\sin^{2}2\theta_{13}$ is 
similar to that of $^7$Li.
The $^{11}$B yield ratio is about 1.27 even in the case of adiabatic 
range of $\theta_{13}$ and normal hierarchy.
This value is much smaller than that of $^7$Li.
Neutrino oscillations raise the rate of 
$^{12}$C($\nu_{\rm e},{\rm e^-}p)^{11}$C and the $^{11}$C yield.
However, the increased $^{11}$C yield is still 
small for the total yield of $^{11}$B.
In the inverted hierarchy case, the maximum yield ratio of $^{11}$B is 
slightly smaller than that in the normal hierarchy.
In this case the contribution of
$^{12}$C($\bar{\nu}_{\rm e},{\rm e^+}n)^{11}$B and
$^4$He($\bar{\nu}_{\rm e},{\rm e^+}n)^3$H increases.
As shown in $^7$Li case, however, the increment is small
due to a small difference of the average energy between
$\bar{\nu}_{\rm e}$ and $\bar{\nu}_{\mu,\tau}$.
In the limit of $\sin^{2}2\theta_{13}=0$, the $^7$Li and $^{11}$B yields 
still slightly increase due to the residual mixing other than 13-mixing
as shown in Fig. 2.

%We note that we incorporate the effect on neutrino oscillations into
%only six charged-current $\nu$-process reactions of $^4$He and $^{12}$C.
%We use the rates without oscillations 
%for the other charged-current $\nu$-process reactions.
%There are two important roles for the $\nu$-process:
%one is direct production through the $\nu$-process and the other
%is indirect; capture reactions by protons and neutrons produced through the
%$\nu$-process.
%For the former, we concentrate on the production of $^7$Li and $^{11}$B,
%both of which are mainly produced from $^4$He and $^{12}$C.
%The production of the other species through the $\nu$-process
%is beyond the scope of our present study.
%For the latter, almost all protons and neutrons are produced from
%$^4$He in the He layer and the contribution from other species through
%the $\nu$-process is negligible.

We solve for neutrino energy spectra changed by neutrino oscillations in
the density profile of a presupernova star.
%, although we calculate
%explosive nucleosynthesis considering the shock propagation.
We expect that the influence on the spectral changes due to neutrino 
oscillations caused by the passing shock is small.
%In O-rich layers, the oscillation amplitude is too small to change
%the amount of $^7$Li and $^{11}$B owing to neutrino oscillations 
%because of high density.
When the shock front is in the O-rich layers, the density behind the shock
front is still so high that the shock wave does not affect the oscillations.
After the shock front passes through the O/C layer,
the change of the density profile affects the mixing probability of 
neutrino oscillations.
However, most neutrinos have passed before the shock arrival at the O/C
layer.
Details are discussed in \cite{yk052}.
%However, the shock arrival time at the O/C layer is about 5 s whereas
%the decay time of the neutrino flux is 3 s.
%More than 80\% of neutrinos have passed through the
%supernova ejecta by the time the shock front reaches the O/C layer.
%Only a small fraction of neutrinos are affected by neutrino oscillations
%in layers altered by the shock.

%The effect of neutrino oscillations on the $^7$Li and $^{11}$B yields 
%also depends on the stellar mass.
%The region of the O-rich layer increases with stellar mass 
%(e.g., \cite{nh88}).
%Since $^{11}$B are produced in the O-rich layers and the He/C layer,
%the contribution from the O-rich layers increases with the stellar mass.
%The O-rich layers are not affected by neutrino oscillations.
%Therefore, the dependence of the $^{11}$B yield on $\sin^{2}2\theta_{13}$ 
%would become smaller as the mass of a star is larger.
%The density structure of the He layer may affect the dependence of the
%$^7$Li yield.
%In our study, the $^7$Be yield is larger by a factor of at most 3.1 
%owing to neutrino oscillations.
%Thus, a stellar model in which $^7$Li is originally produced as $^7$Be 
%rather than $^7$Li may produce a total $^7$Li yield of more than twice 
%the yield of a model without oscillations. 

\begin{figure}[t]
\includegraphics[width=7cm]{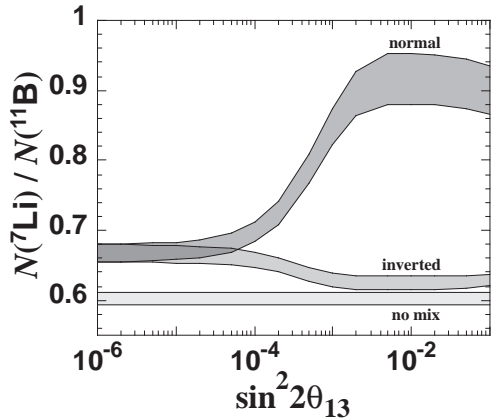}
\caption{
The number ratio of $^7$Li/$^{11}$B with the relation of 
$\sin^{2}2\theta_{13}$.
The shaded ranges include the uncertainties of neutrino energy spectra
deduced from the calculations using three sets of neutrino temperatures
and total neutrino energies (see text).
}
\end{figure}

In our previous studies \cite{yt04,yk05}, we constrained the neutrino 
temperature with Galactic chemical evolution (GCE) arguments 
(e.g.,\cite{fo00}) 
for $^{11}$B. However, the remaining model uncertainties still render the 
effects on the observed $^7$Li and $^{11}$B abundance trends from neutrino 
oscillations somewhat ambiguous. Still, the possibility to obtain, or at least
constrain, fundamental neutrino properties from these observations encourage 
us to pursue these arguments further. We consider the dependence of the 
$^7$Li/$^{11}$B ratio on $\sin^{2}2\theta_{13}$ taking account uncertainties 
of neutrino energy spectra. We consider two additional spectral parameter sets:
$(T_{\nu_{\rm e}}, T_{\bar{\nu}_{\rm e}}, T_{\nu_{\mu,\tau}}, E_\nu)$ =
(3.2 MeV, 5 MeV, 6.6 MeV, $2.4 \times 10^{53}$ erg) and 
(3.2 MeV, 4.3 MeV, 5.2 MeV, $3.5 \times 10^{53}$ erg).
The obtained $^{11}$B yields for these two cases without neutrino oscillations
are $7.3 \times 10^{-7} M_\odot$ and $3.3 \times 10^{-7} M_\odot$, 
corresponding to the maximum and minimum values satisfying the GCE models 
for $^{11}$B \cite{yk05}.
The corresponding $^7$Li yields are $2.9 \times 10^{-7} M_\odot$ and
$1.3 \times 10^{-7} M_\odot$.

Figure 3 shows the number ratio of $^7$Li/$^{11}$B as a function of
$\sin^{2}2\theta_{13}$.
The uncertainty due to neutrino spectra is included as shaded regions.
The $^7$Li/$^{11}$B ratio in the case of adiabatic 13-mixing resonance
and normal hierarchy is larger than that without neutrino oscillations,
even with the spectral uncertainties included.
Thus, the enhancement of observed $^7$Li/$^{11}$B ratio may constrain
the lowest value of $\theta_{13}$ and eliminate the possibility of
inverted hierarchy.
We should note that uncertainties in the $\nu$-process cross sections still
remain.
We expect, however, that they are largely canceled out when we take the
$^7$Li/$^{11}$B ratio.
Since $^7$Li and $^{11}$B are mainly produced through the $\nu$-process
from $^4$He, the dependence of their yields on the $\nu$-process reaction 
rates is similar.
In addition, the dependence of neutral-current reaction rates on the
neutrino temperature is not so different from that of the corresponding
charged-current reactions.
Data analysis of SN 1987A \cite{mr05} and future observations of SN 
relic neutrinos \cite{sk03} may provide additional information on the  
$\bar{\nu}_{\rm e}$ spectrum. The effect of neutrino oscillations on the 
analyzed $\bar{\nu}_{\rm e}$ signal should be taken into account, and the 
evaluation of the $\bar{\nu}_{\rm e}$ spectrum will lead to a more precise 
evaluation of $^7$Li/$^{11}$B ratio in SNe. 
%We note, however, that significant uncertainties in the $\nu$-process cross 
%sections still remain.

Recent observational efforts to obtain Li and B abundances in stars which 
may have formed in regions directly affected by prior generations of massive 
stars and their subsequent SNe (e.g., \cite{pd98}), may have detected 
the signature of the $\nu$-process in $^{11}$B-enriched stars \cite{rp00}.
The combination of SN nucleosynthesis theory and observations of light
elements may ultimately provide powerful constraints on mass hierarchy and 
the mixing angle $\theta_{13}$.

In summary, we investigated light element synthesis in SNe through the
$\nu$-process including the change of neutrino spectra due to neutrino 
oscillations. 
In the case of adiabatic 13-mixing resonance and a normal 
hierarchy, the $^7$Li yield increases by about a factor 1.9 compared to 
the case without oscillations. 
This increase may be accessible to high resolution spectroscopic studies of 
stars in young, star-forming regions.
The $^7$Li yield in other cases and the $^{11}$B yield are scarcely
affected by neutrino oscillations.
The adiabaticity of the 13-mixing resonance and the mass hierarchy affect
robust determinations of $^7$Li/$^{11}$B ratios in SNe.

%\clearpage 

%\begin{acknowledgments}
We would like to thank Koichi Iwamoto, Ken'ichi Nomoto, and Toshikazu 
Shigeyama for providing the data for the internal structure of progenitor 
model 14E1 and for helpful discussions.
Numerical computations were in part carried out on general common use 
computer system at Astronomical Data Analysis Center, ADAC, 
of National Astronomical Observatory of Japan.
This work has been supported in part 
%by the 21st Century COE Program ^^ ^^ Exploring New Science by Bridging 
%Particle-Matter Hierarchy'' in Graduate School of Science, Tohoku University, 
by the Ministry of Education, Culture, Sports, Science and Technology, 
Grants-in-Aid for Young Scientist (B) (17740130) and Scientific Research 
(17540275), for Specially Promoted Research (13002001),
and by Mitsubishi Foundation.
%\end{acknowledgments}

%\bibliographystyle{apsrev}
%\bibliography{yoshidafile}

\end{document}